\documentclass[12pt,epsfig,amsfonts]{article}
\usepackage{graphicx}
\usepackage{color}
\def\be{\begin{equation}}
\def\ee{\end{equation}}
\newcommand{\ba}{\begin{eqnarray}}
\newcommand{\ea}{\end{eqnarray}}
\newcommand{\baa}{\begin{eqnarray*}}
\newcommand{\eaa}{\end{eqnarray*}}
\newcommand{\bb}{\begin{thebibliography}}
\newcommand{\eb}{\end{thebibliography}}

\newcommand{\beq}{\begin{equation}}
\newcommand{\eeq}{\end{equation}}
\newcommand{\beqs}{\begin{eqnarray}}
\newcommand{\eeqs}{\end{eqnarray}}

\begin{document}

\parskip=0.3cm

%\begin{titlepage}

%\vskip 0.5cm \centerline{\Large \bf
%Oscillation of the scattering amplitude    \\ and
%Extra dimension
%and a Gravitational Potential  \\  with taking into account the nucleon form-factor.
%} \vskip 0.2cm  \vskip 0.3cm \centerline{
% O. Selyugin $^{1}$}

\title{Gravitation interaction with extra dimension and
periodic structure  of the hadron scattering amplitude
}
\date{ }

\maketitle

\vskip 0.2cm

\centerline {\sl BLTP, Joint Institute for Nuclear
Research,} \centerline {\sl  141980 Dubna, Moscow region, Russia}

\vskip 0.2cm

\begin{abstract}
  The behavior of the hadron scattering amplitude
  determined by the
  gravitation interaction of hadron at high energies
  with impact of the KK-modes in d-brane models of gravity is examined.
  The possible periodic structure of the scattering amplitude
  and its dependence on the number of additional dimensions
  are analyzed.
  The effects of the  gravitational hadron form factors
  obtained from the hadron generalized parton distributions (GPDs)
  on the behavior of the interaction potential and the scattering amplitude are  analyzed.
  It is shown  that in most part
  the periodic structure comes  from the approximation of our calculations.
\end{abstract}

\vskip 0.1cm
%\keywords{high energy \and hadron \and gravitaion \and n-dimentional
%   \and spin correlation}
% \PACS{PACS 04.30.-w \and PACS 04.80.Cc
%\and PACS 04.50.+h \and PACS 04.60.-m \and PACS 13.85.-t
%\and PACS 13.85.+e }
%\vfill

$
\begin{array}{ll}
^{1}\mbox{{\it e-mail address:}} &
\mbox{selugin@theor.jinr.ru}
\end{array}
$

%\newpage
\section{Introduction}	

Research of a structure of  elastic
  hadron scattering amplitude
   at super-high energies and small momentum transfer - $t$
%   is a very important problem as in this area we should find
  can gives a connection   between
   an experimental knowledge and  the basic asymptotic theorems,
   which are based on the first principles. % \cite{fp}.
   It gives  an information about  hadron interaction
   at large distances where the perturbative QCD does not work,
   and a new theory, as, for example, instantons or string theories,
    must be developed.
   In the early 70s there were
   many works in which the consequences of breaking the
   Pomeranchuk theorem were investigated \cite{royt}.
   It was shown  \cite{akm} that
   if the Pomeranchuk
   theorem is broken and the scattering amplitude
   grows to a maximal possible extant, but not breaks the Froissart boundary,
   many zeros in the scattering amplitude
   should be available in the nearest domain of
   $t \rightarrow 0$ in the limit $s \rightarrow \infty$.
%   So if we
   Hence, with increase energy of colliding beams %at some $s$
%   we should   find this
   some new effect
   in a form  of small oscillations
   in  differential cross sections  can be discovered at very small $t$.

The differential cross section % and analyzing power $A_N$
 are defined as follows:
\begin{eqnarray}
\frac{d\sigma}{dt}&=& \frac{2
\pi}{s^2}(|\Phi_1|^2+|\Phi_2|^2+|\Phi_3|^2
  +|\Phi_4|^2+4|\Phi_5|^2), \label{dsth}.
%  A_N\frac{d\sigma}{dt}&=& -\frac{4\pi}{s^2}
%                 Im[(\Phi_1+\Phi_2+\Phi_3-\Phi_4) \Phi_5^{*})],  \label{anth}
\end{eqnarray}
   Every amplitudes contain the hadronic and electromagnetic parts.
\begin{eqnarray}
   \Phi_i = \Phi_{i}^{em}+\Phi_{i}^{h} \ exp[-i\alpha \varphi]
 \end{eqnarray}
  where
  \begin{eqnarray}
   \varphi(s,t) = \pm \ [\gamma \ + log(B(s,t)|t|/2) \ + \ \nu_1 + \nu_2]
 \end{eqnarray}
   is the Coulomb-hadron interference phase with $ \nu_1 $ connect with taking into account the second Born diagram (2 photon approximation)
   \cite{Sel-phase1} and $\nu_2$ is determined
   by the hadron form factor  \cite{Sel-phase2}.
  At high energy and small angles of scattering it is usually neglect
  the spin-flip amplitudes.
  The scattering at small momentum transfer
  is determined by the interference of the electromagnetic and hadronic parts.

      If there is some unknown additional amplitude with non-small
       real part the differential cross sections is
%       here will be also additional interference term
       \begin{eqnarray}
  d\sigma/dt(s,t) \sim &&|Re F_C(t) +Re F_h(s,t) + ReF_{ad}(s,t)|^2  \\ \nonumber
 && +|Im F_C(t) +Im F_h(s,t) + Im F_{ad}(s,t)|^2,
 \end{eqnarray}
 and main contribution at small momentum transfer is determined by
 the Coulomb-hadron interference term
\begin{eqnarray}
  \Delta(d\sigma/dt)_{ad}(s,t) \sim && [2 Re_C ImF_h (\rho(s,t)+ sin[\alpha_{em} (\varphi_{C}(t) + \varphi_{Ch}(s,t))] \\ \nonumber
  && + 2 Re_{ad}(s,t) [ReF_C(t) + \rho ImF_h (s,t)].
 \end{eqnarray}
where $F_{C} = \mp 2 \alpha G^{2}/|t|$ is the Coulomb amplitude;
$\alpha$ is the fine-structure constant  and $G^{2}(t)$ is  the  proton
electromagnetic form factor squared;
$Re\ F_{N}(s,t)$ and $ Im\ F_{N}(s,t)$ are the real and
imaginary parts of the nuclear amplitude;
$\rho(s,t) = Re\ F(s,t) / Im\ F(s,t)$.

  It is clear that the size and energy dependence of the basic parameters
  of the elastic scattering amplitude at small $t$ are mostly determined by
  a potential of the hadron  interaction at large distances.
   However, as the parameters of these potential in the $r$-representation
   are connected with the real physical effects at small $q$
  by the integral transformation and on the background of the leading
  contributions.  The possibilities of  experimental research
  are limited, though there are some  proposals, for example
  \cite{arb}.

   Standard Regge representation shows that the energy dependence of the
   scattering amplitude depends on the spin of the exchanged $t$-channel particles.
   So the exchange of a graviton of spin  2 leads to a growth
   of the scattering amplitude with energy proportional to $s$ \cite{tHooft1}.
    As the usual gravitational interaction
   is very small, this growth can be seen at the Plank scale. However,
   a modern development of the fundamental theory, introduced long ago \cite{kk1,kk2}, is connected with
 the fruitful idea that spacetime has a dimension higher than $D=4$.

 Now there are many different paths in the development of these ideas
 %. Note, the model
 %of the space-time foam as a gas of D-particles in the bulk space-time of the higher-dimentional world
 \cite{Mavr0}.
 Especially it is connect with the different supersymmetry and string models.
% ADD, PLB436(1998)257,
For example in \cite{ADD-PLB246}, was analized the proposal for using large (TeV) extra dimensions in the
Standard Model, motivated from the problem of supersymmetry breaking in string theory, and \cite{ADD-PLB436},where it gave the string realization of low scale gravity and braneworld models, and pointed out the motivation of TeV strings from the stabilization of mass hierarchy.

%work PLB246(1990)317,

A number of studies of higher-dimensional (Kaluza-Klein) field theories were carried out
\cite{rev,rubakov1,perez,rand1,rand1b}.
In a modern context,
the Kaluza-Klein (KK) theories arise naturally from (super)string theories
in the limit
where the relevant energies $E$ are much smaller than the string mass scale
$M_s \sim (\alpha')^{-1/2}$, $\alpha'$ being the slope parameter.
Since field theories of gravity
behave badly in the ultraviolet limit, Kaluza-Klein formulations should in general be
 regarded as effective actions, with an implicit or explicit ultraviolet
 cutoff $\Lambda$ \cite{han}.
  As a first approximation, we may suppose that all Standard Model fields
 are confined to a four-dimensional brane world-volume.
  In the Arkani-Hamed, Dimopoulos and Dvali approach (ADD) \cite{add}
 a large number $d$ of extra dimensions is responsible for a lower Planck
 scale, down to a TeV and only the graviton propagates in the $4+d$ dimensions.
 This propagation manifests itself in the standard $4$ dimensions as a tower of massive
 KK-modes. The effective coupling is obtained after summing over all the KK
 modes and, due to the high multiplicity of the KK modes, the effective interaction
has strength  $1/M_d$ \cite{nussinov,wells}.
Setting $M_{4+d}^{d+2}=(2\pi)^d \hat M_{4+d}^{d+2}$, as in
Ref. \cite{add} (motivated by toroidal compactification, in which the volume
of the compactified space is $V_d = (2\pi r_d)^d$) and applying Gauss's law at
$r<< r_d$ and $r >> r_d$, one finds that
$M_{Pl}^2 = r_d^d M_{4+d}^{2+d} \ \ \ $.
%so that
%$$  \ \ r_d =  \Biggl ( \frac{M_{Pl}}{M_{4+d}} \Biggr )^{2/d}/ M_{4+d} .$$
   In the higher-dimensional models with a warped extra dimension \cite{rand1,rand1b},
  the first KK mode of the graviton can have a mass of the order of $1 \ $TeV
 and the coupling with matter on the visible brane
  is of the order $1 \ $TeV$^{-1}$.
    There are also "intermediate" models  with a small warp  which
  consider the brane  as almost flat. % \cite{sm-war}.
  Such models remove some cosmological bounds on the number of additional
  dimensions.
    All these models provide some experimental possibilities to check (or discover)
  the impact of the extra dimensions on our $4$-dimensional world.
  Now in many papers  new effects are examined which in principle can be seen
  at future colliders.
   In \cite{FPhys}, we show that these effects can also be discovered in
  experiments on elastic polarized hadron scattering.
  We explore in this case the sensitivity of interference spin effects to
  small corrections to the scattering amplitude, linear rather than quadratic
  (as  in the case of cross sections) functions of a small parameter.
%  This was property was first pronounced in the phenomenon of Coulomb-Nuclear
%  interference (CNI) \cite{KopLap}

  In this paper we examined the behavior of the gravitation amplitude
  of the hadron hadron scattering in the case of the one and two
  additional dimensions with takin into account the Kaluza-Kline states.
  At high energies such amplitude can give the additional real
  part to the standard hadron scattering amplitude. Especially
  we examine the possible oscillation in this additional part of the
  scattering amplitude. It can be important effect as it can be
  reveal in the future experimental data at LHC.

\section{The graviton contribution with KK-modes}

Assuming that the higher-dimensional theory at short distances is
a string theory, one expects that the fundamental string scale $M_s$ and the Planck
mass $M_{4+d}$ are not too different (a perturbative expectation is that
$M_s \sim g_s M_{4+d}$).
   As of now, only known framework that allows a self-consistent description
   of quantum gravity is string theory \cite{polch-98}.
 Thus, a
compactification radius $r_d << M_{Pl}^{-1}$ corresponds to a
short-distance Planck scale and string mass $M_s$ which are  $<< M_{Pl}$.
So, we apply constrain on the quantum gravity scale $M_D$.
%  the  mass $KK$
  to the  string scale $M_s$.

Following \cite{nussinov,wells},
 the amplitude  taking into account the KK-modes can be written as

\beq
   A_{grav.}     \sim
\int_{0}^{\infty} \ \frac{d^{d-1} q_T}{ q^2 \ + \
q_T^2}  \nonumber \\           =
 \frac{\pi}{2}\frac{1}{q^2}(\frac{1}{q^2})^{-d/2} \csc (d \frac{\pi}{2})
\eeq
This solution has the poles at $d = 2,4,6...$.
However such answer corresponds the bound of $d < 2$.
%In this case we have non-divergent answer.
  In the case of d=1 the scattering amplitude have the $1/q$ behavior.
 When $ 2 \leq d$ the integral divergent. If we take upper bound of the
 integral - $M_s$ we obtain
\beqs
   A_{grav.} &\sim&
\int_{0}^{M_s} \frac{d^{d-1} q_T}{ q^2 \ + \
%-q_0^2 + |{\bf q}|^2 +
q_T^2}  \nonumber \\
&=& \frac{M_s^d}{d} \frac{1}{q^2} {}_{2}F_{1} [1, d /2, 1 + d /2, -M_s^2 m^2/q^2]).
\label{agqb}
\eeqs
  The hypergeometric function $_{2}F_{1}$ has the smooth behavior
  but the upper integral limit income as multiple coefficient
    which leads to divergence of the Born amplitude
 if the upper limit of integral   $ \rightarrow \infty $ and $ 2 \leq d$
%Next, substituting eq. (\ref{kkeffect}) in eq. (\ref{gravamp}), using the
%above approximations, and inserting eq. (\ref{rnvalue}) for $r_n$, we find
%for $n =2$
   For more high additional dimensions we obtain from eq.(7):
\beq
\tilde A_{grav.}^{Born} =
 \sum_{\ell_1, ..., \ell_n} A_{1g+ KK} =
\frac{\pi s^2}{M_{4+2}^4}\ln \Bigl ( 1 + \frac{M_s^2}{q^2} \Bigr )
\label{born}
\eeq
and for $n = 3$,

\beq
\tilde A_{grav.}^{Born} =
%\sum_{\ell_1, ..., \ell_n} A_{1g+ KK} =
\frac{S_d s^2}{M_{4+3}^{4}} \biggl ( \frac{M_s}{M_{4+3}} \biggr )^{n-2}
 [ \ 1 \ - \ \frac{q}{M_s} \ ArcTh (\frac{M_s}{q}) ];
\label{atildengt3}
\eeq
and for $n=4$:
\beq
\tilde A_{grav.}^{Born} =
%\sum_{\ell_1, ..., \ell_n} A_{1g+ KK} =
\frac{S_d s^2}{M_{4+4}^{4}} \biggl ( \frac{M_s}{M_{4+4}} \biggr )^{2}
 [ \ 1 \ - \ \frac{q^2}{M_s^2} \ Ln (\frac{M_s^2}{q^2}) ];
\label{atildengt4}
\eeq
where $S_d$
\beq
S_d= \frac{2\pi^{d/2}}{\Gamma(d/2)}
\label{sd}
\eeq
is the area of the unit sphere in
${ R}^d$.
 In this work we bound himself by the case $d=2$.
The higher dimensional lead to some additional $q$ dependence of amplitude,
but with the some problems of the excluded the large constant term.
which, according \cite{wells}, can be removed.

    If our particles live on the 3-dimensional brane,
   we can obtain the amplitude in an impact parameter representation,
  where the Born amplitude
  corresponds to the eikonal of the scattering amplitude \cite{Amati}
\begin{eqnarray}
  \chi(s,b) \ = \frac{1}{2 \pi} \int_{0}^{\infty} b \
  J_{0}(q b) \ A^{Born}(q^2) \ db.
  \label{born}
\end{eqnarray}

 An exact calculation of (\ref{born}) for $d=2$ gives (see fig.1a)
\begin{eqnarray}
  \chi(b) = \frac{s}{M_{D}} \ (1 - b  M_D \  K_{1}(b \ M_D))/b^2.
\label{chi-our}
\end{eqnarray}
  Some detailed calculations can be found  in our work \cite{st-pr05}.

\section{Oscillations of the amplitude}

 The Coulomb  force, falling as $1/r^2$ lead to the scattering amplitude
  is proportional $1/t$, where the $t$ is the momentum transfer.
  In the second Born approximation there appear the additional phase
  and the amplitude will be
  \begin{eqnarray}
F(s,t)=\frac{2\alpha }{-t} \ e^{i\alpha_e \varphi_C},
 \end{eqnarray}
where $\alpha_e$ is electromagnetic constant, and $\varphi_C = Log(\lambda/q^2)$ withe $\lambda$ effective photon mass.
 This phase divergence when we put the effective photon mass equal zero.
 Obviously, this phase do not impact on the differential cross sections.
 In the case of the hadron electromagnetic interaction it is need  take into account the hadron form-factor and the phase will be
 more complicated \cite{Sel-phase1}
  \begin{eqnarray}
\varphi(s,t)=\varphi_C(t) + \varphi(s,t),
 \end{eqnarray}
 The $\varphi(s,t)$ was calculated in \cite{Sel-phase2}.
 If  take $\alpha_e \sim 1$ and there is some small photon mass such amplitude will
  be oscillate at small momentum transfer (Fig.1a). Of course, the real
    situation leads to the  disappear such periodic structure - $Fig.1b$.

% Fig.1
\begin{figure}
\begin{flushleft}
 \includegraphics[width=0.5\textwidth]{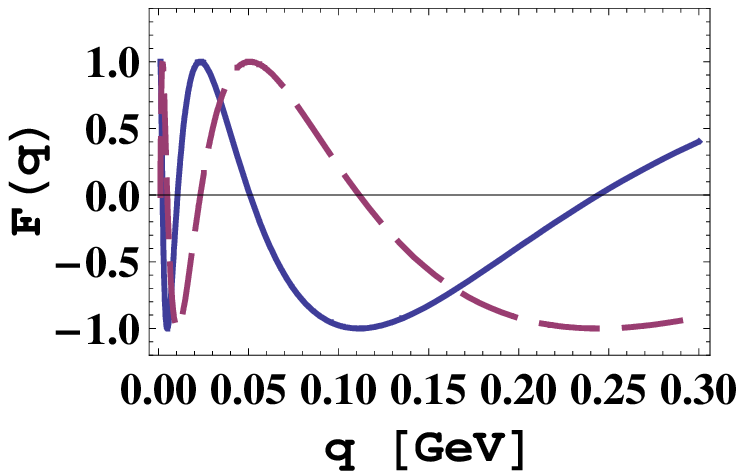}
%\mbox{\epsfysize=50mm\epsffile{gr4chb.ps}}
\end{flushleft}
\vspace{-6.5cm}
\begin{flushright}%\vspace{-7.5cm}
 \includegraphics[width=0.5\textwidth]{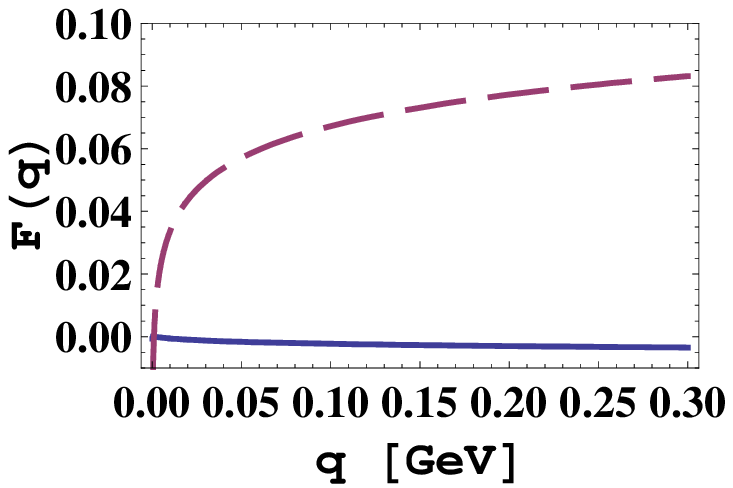}
%\mbox{\epsfysize=50mm\epsffile{v2ll.ps}}
\end{flushright}
\caption{a) [left]
  $F(q)=-F_C(q)*q^2$ with artificial $\alpha_e=1$ and $\lambda=10^{-5}$;
  b) [right] $F(q)=-F_C(q)*q^2$ with real $\alpha_e$ and $\lambda=10^{-5}$; .
  }
\label{Fig_1}
\end{figure}

The Veniziano \cite{ACV} amplitude was obtained also for gravitation force
   but with taking into account the sum of the diagrams using the
   standard eikonal representation
   \begin{eqnarray}
F(s,t)=\frac{8 \pi \alpha_G}{q^2}(\frac{4}{q^2})^{-i\alpha_G},
 \end{eqnarray}
 where $\alpha_G = Gs$. The additional phase is practically the same as in
 the t'Hoft amplitude.  If $\alpha_G = 5$ there also will be some osciilations at small momentum transfer (Fig.2a).
  However already $\alpha_G=1$ such oscillation disappear - Fig.2b.

%Fig.2
\begin{figure}
\begin{flushleft}
 \includegraphics[width=0.45\textwidth]{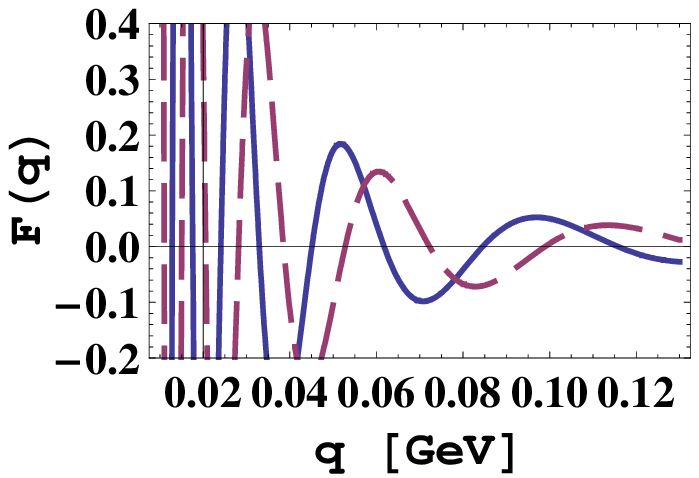}
%\mbox{\epsfysize=50mm\epsffile{gr4chb.ps}}
\end{flushleft}
\vspace{-6.3cm}
\begin{flushright}%\vspace{-3.5cm}
 \includegraphics[width=0.45\textwidth]{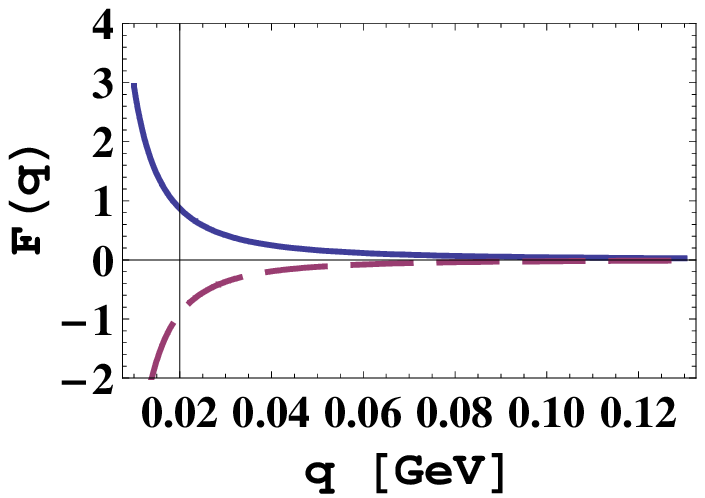}
%\mbox{\epsfysize=50mm\epsffile{v2ll.ps}}
\end{flushright}

\caption{a) [left]
  t'Hoft amplitude $\alpha_G =5$
  b) [right] $\alpha_G =0.1$.
  }
\label{Fig_2}
\end{figure}

 The standard t'Hoft pole \cite{tHooft1} and also \cite{ACV-Pl,MS-PRD}, obtained for the gravitation
    force,
\begin{eqnarray}
F(s,t)=\frac{\Gamma(1-iGs)}{4\pi \Gamma(iGs)}(\frac{4}{-t})^{1-iGs}.
 \end{eqnarray}
 with $G$ - the standard gravitation constant. It
 also contain the additional phase but it leads in this case to the some
  oscillation of the amplitude (Fig.2b) and also disappear in the differential cross sections.

 Taking into account  the additional dimension leads to the renormalization of the gravitation constant and the effective
    gravitation constant $\alpha_G$ can be near the unity or above already
     at LHC energies. The scattering amplitude in the case for
     $n=4+2$ after the eikonalization has also the oscillation behavior
     \cite{wells} and \cite{Aref}. % (Fig.3a,b).
     In that cases the eikonalized amplitude calculated by some
     approximation using the  method of stationary point.
     In the case of the eikonal phase (\ref{chi-our})  for $n=4+2$ gave the amplitude which can be have some periodic structure.
    If we will be calculate the amplitude by numerically it is need
    carefully check up the result, which can be dependence from the
    accurecity and the integral limits. In the standard hadron
    scattering with Gaussian potentials there is not such problems
     as the Gaussian form of phase has good behavior at zero impact parameter
     and faster decreasing at large $b$.
     Let us see how the result will be dependence from the upper bound of
     integral. In the case of the one addition dimension and take the integral in the form as in \cite{Aref}

    \begin{eqnarray}
  A(s,b)_{eik} \ = 4 \pi s b_{0}^{2} F_{n}(b_n q),
  \label{born}
\end{eqnarray}
\begin{eqnarray}
  F_{n}(y) \ = -i \int_{0}^{A} b \
  J_{0}(q b) \ (e^{ix^{-n}} - 1).
  \label{born}
\end{eqnarray}
where input $x=b/b_c$  and take \cite{wells,Aref}
\begin{eqnarray}
 \chi(b)=(b_c/b)^n, \ \ \ {\rm with} b_c=[\frac{(4\pi)^{n/2-1} s \Gamma(n/2)}{2 M_{D}^{n+2}}]^{1/n}
  \label{chi-weis}
\end{eqnarray}
  In this case for $n=1$ the amplitude will be has the same periodic
   structure as in (\cite{Aref} (Fig.3)) if the upper bound of the integral
    $B=25$ (Fig.3a -upper). The period of this oscillation will be
    inverse proportional to the size of upper integral value $A$.
    If the $A=50$ the period decrease in to time.
   If the upper bound of the integral sufficiantly large $a=500$
    the visible oscillations  is disappear (Fig.3b - upper). However,  if it is take a small
    interval ind made the magnification it then the periodic structure  appear again (Fig3.b - low).
    In most part such effect connect with the slowly decreasing the
    eikonal phase in the case $n=1$.
   Such calculation for the $n=2$, using the phase (\ref{chi-our})
   show the practically disappear the periodic structure.

%Fig.3
\begin{figure}
\begin{flushleft}
 \includegraphics[width=0.45\textwidth]{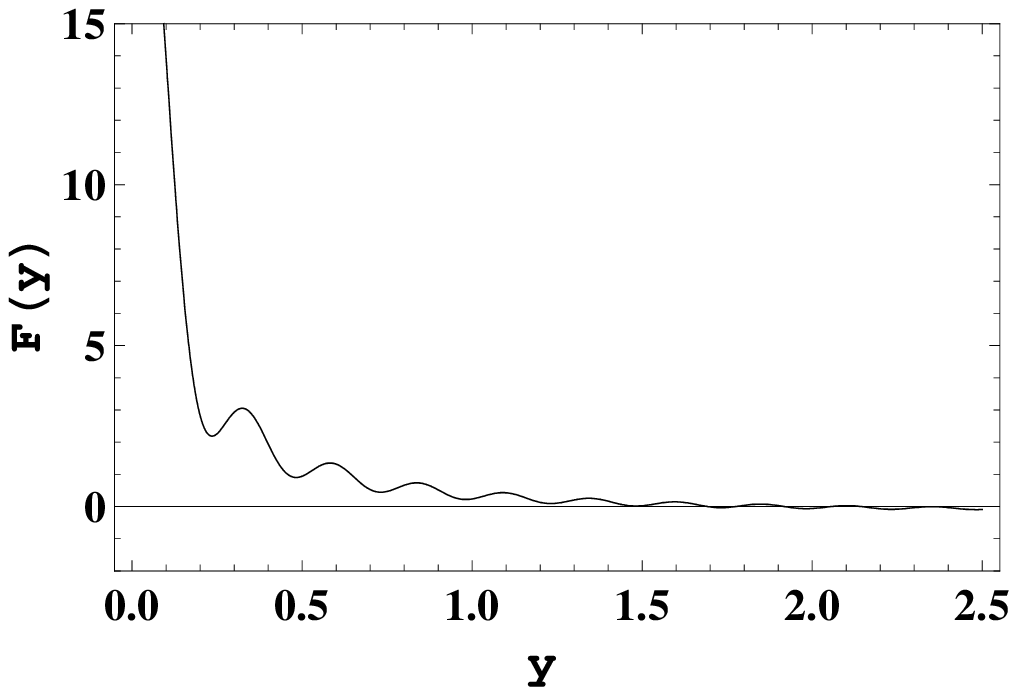}
%\mbox{\epsfysize=50mm\epsffile{gr4chb.ps}}
\end{flushleft}
\vspace{-6.cm}
\begin{flushright}%\vspace{-4.5cm}
 \includegraphics[width=0.45\textwidth]{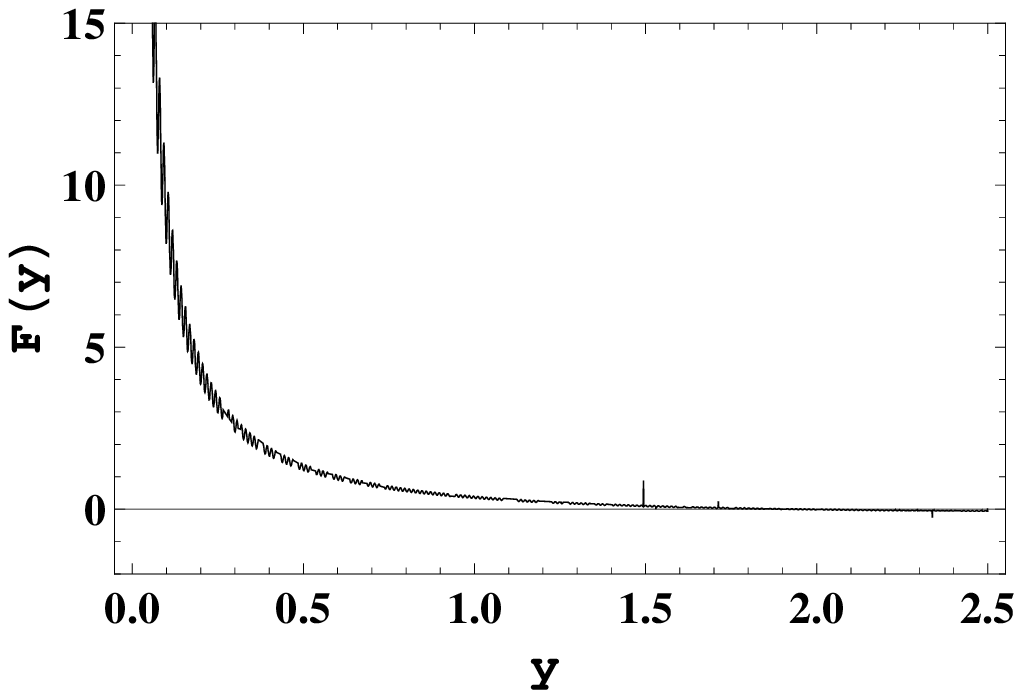}
%\mbox{\epsfysize=50mm\epsffile{v2ll.ps}}
\end{flushright}

%\phantom{.}
%\vspace{0.5cm}

\begin{flushleft}
 \includegraphics[width=0.45\textwidth]{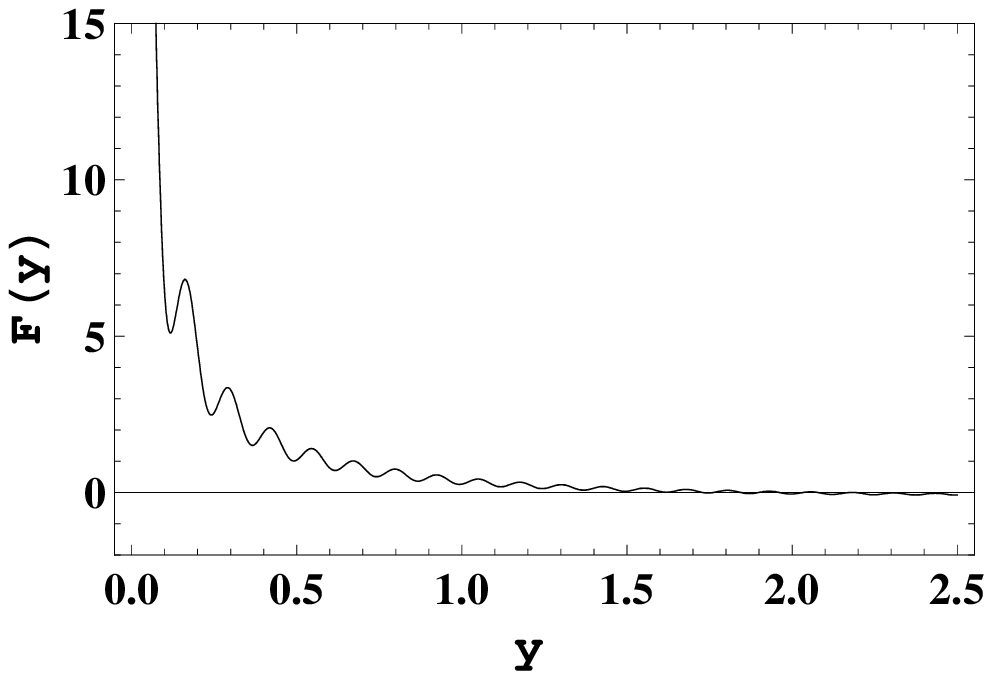}
%\mbox{\epsfysize=50mm\epsffile{gr4chb.ps}}
\end{flushleft}
\vspace{-6.1cm}
\begin{flushright}%\vspace{-4.5cm}
 \includegraphics[width=0.45\textwidth]{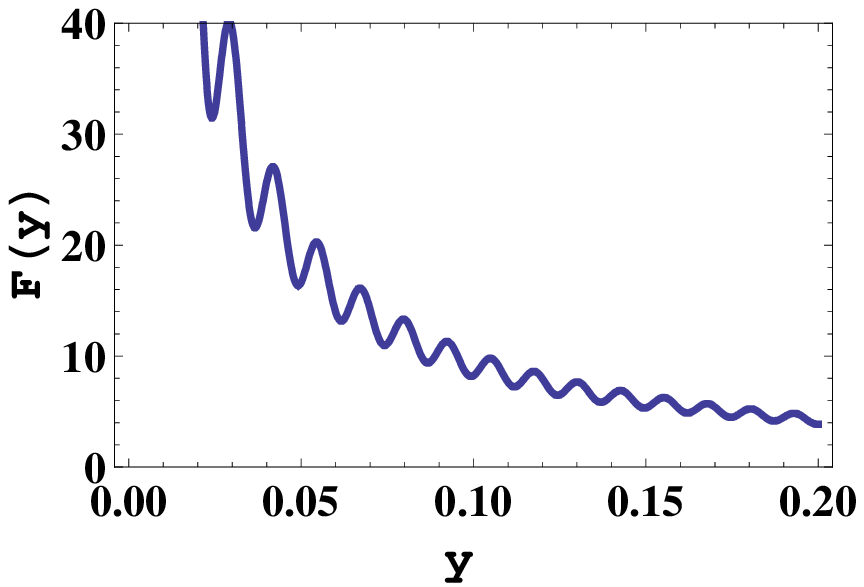}
%\mbox{\epsfysize=50mm\epsffile{v2ll.ps}}
\end{flushright}
\caption{a) [left]
  Dependence over upper integral bound
  a)[left] upper - $B=25$; low - $B=50$;
  b) [right] upper -$B=500$; low - $B=500$ ( magnification)
  }
\label{Fig_1}
\end{figure}

\section{Conclusion}
 Taking the cut of the integration of the contribution of the
  the Kaluza-Kline states, which is proportional string mass $M_s$ the gravitation Born amplitude in the n-dimensions was obtained. The corresponding eikonal phase for the two additional dimension show the change the behavior of phase at small impact parameter from the standard
  $1/b^2$ behavior.
  The numerical calculation of the eikonalized amplitude show
   the some periodic structure in the case of one additional dimension.
   However such structure has the period which is inverse proportional to the upper bound of the eikonal integration. It means that if the such gravitation potential has some saturation regime at large, but not huge,
   distances (for example, order $50 fm$) it can leads to the some small
   oscillation behavior of the diffraction differential cross sections.
   Of course, in the real case we need take into account the
    form factors of the scattering particles.

     Our calculations of the impact of two additional dimensions  on the gravitational potential      at small distances  show that it differs from  a simple power
     % dependence of the gravitational potential
      at $r \leq 3 \ $GeV$^{-1}$ and it changes
     the  profile function of  proton-proton scattering.
     Including  the effects of Kaluza-Klein modes of graviton scattering   amplitudes, with two extra dimensions and taking into account  the gravitational form factor, which we calculated from the GPDs of the nucleons $A(x,t)$, it was shown  that   the impact-parameter dependence of the gravitational eikonal   heavily changes from the standard $1/b^2$  dependence. This is the main result of our paper.
%     Taking into account the gravitational form factor, which we calculated
%     from the GPDs of the nucleons $A(x,t)$, leads to an essential change
%     of the impact-parameter
%     dependence of the gravitational eikonal.
%     Its form heavily changes from the standard $1/b^2$ dependence. This is the main result of our paper.
     We think that this effect has to be taken into account in the calculation of the production
     of Black Holes  at  super-high energy accelerators.

     We have  shown that the gravitational interaction  additional dimensions and
     the  possible small spin-flip amplitude, proportional to the gravitational
     form factor $B(t)$, lead to large spin correlation effects at small angles and
      $-t \sim 10-30 \ $GeV$^2$. However, the inclusion of  the gravitational form factors
      $A(b)$ decreases this effect  and drastically changes its form.

\section*{Acknowledgments}
The authors express their thanks to Ir. Arefieva and O. Teryaev
for fruitful discussions.

\end{document}